\documentclass[aps,pra,twocolumn,notitlepage,superscriptaddress,showpacs,10pt,longbibliography]{revtex4-2}
\usepackage{times}
\usepackage{amsmath}
\usepackage{amssymb}
\usepackage{txfonts}
\usepackage[table]{xcolor}
\usepackage{hyperref}
\usepackage{color}
\usepackage[utf8]{inputenc}
\usepackage{amstext}
\usepackage{physics}
\usepackage{ulem}
\makeatletter
\let\openbox\@undefined
\makeatother
\usepackage{amsthm}
\usepackage{appendix}
\usepackage{xr}
\makeatletter
\theoremstyle{plain}

\theoremstyle{remark}

\usepackage{amsmath,amssymb,bm,physics,mathtools}
\usepackage{siunitx}
\usepackage{graphicx}
\usepackage{microtype}
\usepackage{enumitem}
\usepackage{booktabs}
\usepackage{float}
\usepackage{hyperref}
\setlist{nosep}
\usepackage{ulem}
\usepackage{accents}

\newcommand{\rvec}{\mathbf{r}}
\newcommand{\pvec}{\mathbf{p}}
\newcommand{\uvec}{\mathbf{u}}
\newcommand{\fvec}{\mathbf{f}}
\newcommand{\Vvec}{\mathbf{V}}
\newcommand{\Fvec}{\mathbf{F}}
\newcommand{\tvec}{\boldsymbol{\tau}}
\newcommand{\wvec}{\boldsymbol{\Omega}}

\newcommand{\dvec}{\mathbf{d}}

\newcommand{\uu}{\smash[b]{\uline{\hbox{u}}}}
\newcommand{\ff}{\smash[b]{\uline{\hbox{f}}}}
\newcommand{\Omat}{\smash[b]{\uuline{\hbox{O}}}}
\newcommand{\Rotmat}{\smash[b]{\uuline{\hbox{R}}}}
\newcommand{\Imat}{\smash[b]{\uuline{\hbox{I}}}}
\newcommand{\Gmat}{\smash[b]{\uuline{\hbox{G}}}}
\newcommand{\Amat}{\smash[b]{\uuline{\hbox{A}}}}
\newcommand{\Bmat}{\smash[b]{\uuline{\hbox{B}}}}
\newcommand{\Cmat}{\smash[b]{\uuline{\hbox{C}}}}

\newcommand{\Mmat}{\smash[b]{\uuline{\mathbf{M}}}}

\newcommand{\Qmat}{\smash[b]{\uuline{\mathbf{Q}}}}
\newcommand{\Pmat}{\smash[b]{\uuline{\mathbf{P}}}}

\newcommand{\notequal}{\neq}

\newcommand{\wc}[1] {{\color{black}#1}}

\newcommand{\hd}[1] {{\color{black}#1}}
\newcommand{\jb}[1] {{\color{black}#1}}
\makeatother

\begin{document}

\title{STL-to-Stokeslet Computation of Mobility Tensors and Sedimentation Dynamics for Shaped Particles}

\author{Wenting Cheng}
\affiliation{Department of Physics, Emory University, Atlanta, GA 30322, USA}

\author{Tiago Pernambuco}
\affiliation{Department of Theoretical and Experimental Physics, Federal University of Rio Grande do Norte, 59078-970, Natal, Brazil}

\author{Thomas A. Witten}
\affiliation{Department of Physics and the James Franck Institute, The University of Chicago, Chicago, IL 60637 USA}

\author{Haim Diamant}
\affiliation{School of Chemistry, Tel Aviv University, Ramat Aviv, Tel Aviv 69978, Israel}

\author{Justin C. Burton}
\affiliation{Department of Physics, Emory University, Atlanta, GA 30322, USA}

\begin{abstract}

Sedimentation is extremely common in nature, occurring throughout the atmosphere and oceans, and in every laboratory centrifuge. The shape and mass distribution of a particle uniquely determines its motion at low Reynolds number, and complex dynamics can emerge from even simple particle shapes.  The dynamics are governed by the particle’s hydrodynamic mobility tensor, which \wc{dictates the translational and rotational velocities given the forces and torques}. However, to date the inference of the mobility tensor from the object shape has been cumbersome and tricky. Starting with an input file representing an object for a 3D printer, such as an STL file, here we present an efficient numerical framework to compute the mobility tensor by discretizing the particle surface into distributed point drag forces called stokeslets. We validate our results against analytical solutions of simple geometries and recent experimental measurements. 
With our calculated mobility tensors in hand, using standard transformation laws, we demonstrate the dramatic effect of shifting the center of mass from a center of symmetry: all initial orientations evolve into \wc{one, two, or three particular final motions dictated by the object.}
By providing a user-friendly and efficient framework to compute the mobility tensor and resulting particle dynamics, this work offers a broadly applicable tool for the soft-matter, fluid-mechanics, and biophysics communities, and facilitates the design of steerable particles under diverse external forces, with relevance to colloidal transport, biological locomotion, diffusion, and self-assembly.

\end{abstract}

\maketitle

Significance statement: Sedimentation is a fundamental process across many natural systems, from atmospheric aerosols and marine snow to the locomotion of microorganisms against gravity.  The shape of a particle crucially determines its sedimentation behavior, often leading to complex rotational and orbital trajectories. However, predicting these dynamics for arbitrary particle shapes remains a challenge. Here we demonstrate an efficient numerical framework to compute sedimentation dynamics in viscous fluids directly from a three-dimensional model. The framework employs a user-friendly computational interface, can handle highly complex particle shapes and density inhomogeneities, and serves as a predictive tool for tuning locomotive properties of particles in viscous fluids and their responses to \wc{various} external forces.\\


The motion of particles in viscous fluids is governed not only by external forces but also by particle shape and mass distribution\wc{; these} can induce nontrivial coupling between translation and rotation~\cite{witten2020review}. A simple example is marine snowfall, where complex particle shapes form through aggregation and sediment \wc{over} years \cite{chajwa2024hidden}. \wc{Moreover}, shape can directly influence interactions, leading to both attractive and repulsive effective forces \cite{goldfriend2017screening,nissanka2023dynamics,joshi2025dynamics}. Yet, the dynamics of even a single shaped particle is rich, exhibiting fixed points, limit cycles, and slow transients manifested as twists, tumbles, and orbital motion. 
Predicting these dynamics requires accurate evaluation of the particle’s \emph{$6\times6$ mobility tensor} ~\cite{kim2013microhydrodynamics,happel2012low,witten2020review,gonzalez2004,krapf2009}, which fully determines how forces and torques are converted into translational and rotational motion \wc{in the creeping-flow regime where inertial effects are negligible}.
Classical analytical results for rigid-body hydrodynamics are largely restricted to highly symmetric shapes such as spheres~\cite{happel2012low}. Numerical methods often represent particles as assemblies of point forces (stokeslets) acting on the fluid~\cite{kirkwood1948intrinsic}, and the resulting flow fields and particle motions can be obtained through linear algebraic operations~\cite{youngren1975stokes}. However, their validation and practical application have remained largely confined to relatively simple geometries such as spheroids and cylinders. This limitation motivates the development of efficient and generalizable approaches capable of treating particles with arbitrary shape.

Modern methods for structure-based hydrodynamic modeling are now well developed~\cite{brookes2025bead}. A popular computational approach treats particles as assemblies of beads with varying degrees of complexity in their interactions \cite{de1981hydrodynamic,de2000calculation,ortega2011prediction,rai2005somo,rocco2015computing,brookes2018recent,rocco2021us,garcia2002hydrodynamic,zuk2018grpy,collins2021lord}. Many of these methods are designed to efficiently predict single-valued transport properties, for which fine geometric features are largely irrelevant~\cite{de2016hydro}. As a result, they do not provide direct access to the mobility tensor. High-accuracy methods based on multipole expansions~\cite{durlofsky1987dynamic,cichocki1999lubrication,ekiel2009precise} or boundary element formulations~\cite{aragon2004precise,aragon2011recent} can in principle recover the exact mobility tensor, however, their substantial computational cost and limited scalability have largely restricted their use to benchmark studies or simplified geometries.
Recent studies have sought to access the mobility tensor using bead-based simulations combined with experimental trajectory fitting~\cite{huseby2025helical}, direct inference from sedimentation experiments~\cite{miara2024dynamics}, or theoretical calculations based on boundary integral methods~\cite{pozrikidis1992boundary,kim2013microhydrodynamics} for highly symmetric particles~\cite{joshi2025sedimentation}. While successful for particular geometries, these approaches remain either computationally intensive or specialized, underscoring the need for a general, efficient framework that directly links arbitrary particle geometry to its full rigid-body hydrodynamic response.

To overcome these limitations, we introduce an \emph{STL-to-stokeslet} framework that computes the rigid-body mobility tensor and the resulting dynamics directly from an STL file of an arbitrarily shaped particle. These files are commonly used for 3D printing, and the method requires no geometric decomposition, symmetry assumptions, or experimental calibration. 
Using the stokeslet discretization implicit in the STL primitives and prescribing independent rigid-body motions, the mobility tensor is efficiently extracted and used to predict responses to external forces and torques. We validate our results using analytical solutions of simple particle geometries~\cite{happel2012low} and recent sedimentation experiments using helical ribbons~\cite{huseby2025helical} and \wc{bent disks (i.e., disks bent out of plane, which gyrate despite being nonchiral, sometimes called ``U-shaped'' disks in the literature)}~\cite{miara2024dynamics}. 
The framework further incorporates translations of the center of mass (CoM) through a simple transformation of the mobility tensor. We find that during sedimentation under gravity, the dynamics of certain particle shapes display a pronounced sensitivity to shifts in the CoM smaller than 1\%. Such density variations are common in natural settings and in industry where particles are fabricated. 

The framework is highly efficient: on a standard personal computer, the mobility matrix for $N \simeq 10^4$ stokeslets can be computed in approximately $1\,\mathrm{min}$. By directly linking geometric shape to hydrodynamic response, our approach also provides a versatile framework to rationally design and control steerable particles under external forces. 
This enables a wide range of applications, from engineering locomotion at low Reynolds number, such as bacterial flagella~\cite{purcell1997efficiency,kim2003macroscopic,djutanta2023decoding,rodenborn2013propulsion}, to predictive modeling of the sedimentation and swimming of natural particulate matter, including plankton~\cite{guasto2012fluid,sengupta2017phytoplankton}. 
More broadly, it offers new opportunities for hydrodynamic trapping and flow-directed assembly of particles in solution~\cite{shenoy2016stokes,georgiev2020universal}, and establishes a geometry-based framework for synthesizing microscale swimmers in applications such as 
water decontamination, chemical mixing, targeted drug delivery, and noninvasive microsurgery~\cite{ying2019radioactive,ren20193d,xie2020bioinspired,dillinger2024steerable}.


\begin{figure*}[!htbp]
\centering
\includegraphics[width=\linewidth]{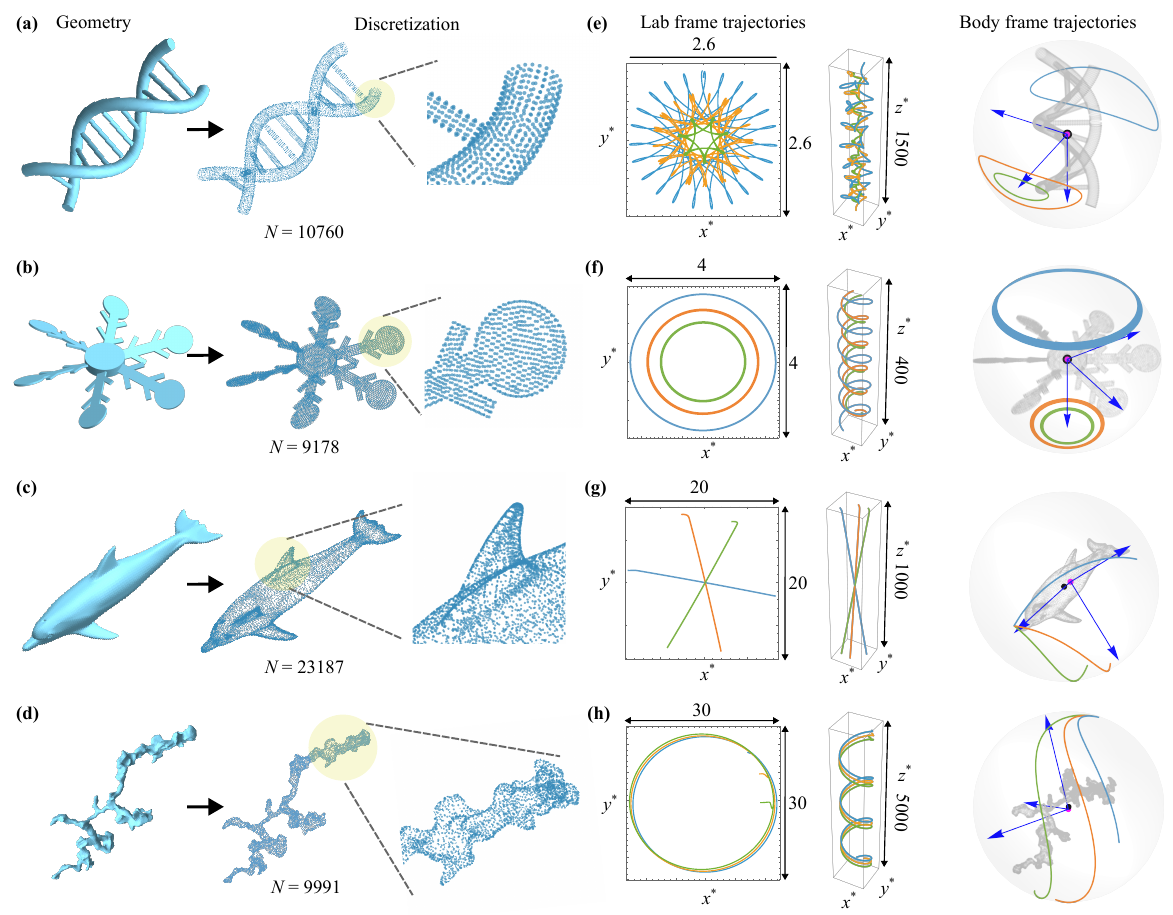}
\caption{
\textbf{
Sedimentation of arbitrarily shaped particles.}
\textbf{(a–d)} Input geometries from STL files (left) and their stokeslet discretizations (middle) for a DNA-shaped helicoid, a chiral snowflake, a dolphin-like body, and a fractal ice grain \cite{nicolov2025dynamics}, respectively. The right panel shows a zoomed-in view of the highlighted region, illustrating the local distribution of the stokeslets on the discretized surface.
\textbf{(e–h)} Predicted CoM sedimentation trajectories under gravity for the four shapes, shown in the $x$--$y$ plane (left), in 3D (middle), and the corresponding force-vector trajectories on the unit sphere (right). Different colors denote different initial orientations. Chiral particles generate helical or spirograph-like trajectories with closed orbits on the unit sphere, while asymmetric shapes such as the dolphin and the ice grain exhibit biased or helical motion whose force-vector trajectories flow to a fixed point.
Magenta points mark the center of reaction (CoR), about which the \wc{$\Bmat$ is symmetric.} 
\wc{Black points mark the CoM. For particles \textbf{(a)} and \textbf{(b)}, the force center and CoM coincide, the $\Bmat$ tensor is symmetric, and the rotation direction follows closed orbits. For particles \textbf{(c)} and \textbf{(d)}, the CoM was defined assuming uniform density, the resulting $\Bmat$ have a large antisymmetric component, and all initial states converge to the same final rotation axis.}
}
\label{fig1}
\end{figure*}

\subsection*{STL-to-Stokeslet method}

We compute the hydrodynamic mobility of an arbitrarily shaped body using a set of $N$ point forces called stokeslets in an otherwise empty quiescent fluid \cite{kirkwood1948intrinsic}. The stokeslet positions $\{\rvec_1,\ldots,\rvec_N \}$ are distributed on the body surface and are assigned using an STL file. Specifically, they are the node positions of a triangulated surface approximating the body shape.  Other content in the STL file is ignored. Since the stokeslets represent a rigid body, they maintain their relative positions as they move through the fluid. Under a given body motion, each stokeslet's velocity $\uvec_n$ is thus imposed. This motion incurs a corresponding drag force $\fvec_n$ on each stokeslet. Stokesian dynamics dictates the velocities for a given set of drag forces.  

Each stokeslet located at $\rvec_n$ generates a velocity field
$\uvec(\rvec)$ in the surrounding fluid given by the Oseen tensor
$\Omat(\pvec)$, where $\pvec=\rvec-\rvec_n$:
\begin{equation}
\Omat(\pvec)\,\fvec
=
\frac{1}{8\pi\mu |\pvec|}
\left(
\fvec + \hat{\pvec}(\hat{\pvec}\cdot\fvec)
\right),
\qquad
\hat{\pvec}=\pvec/|\pvec|,
\end{equation}
\hd{where $\mu$ is the fluid's shear viscosity}. The total velocity field is
$\uvec(\rvec)=\sum_n \Omat(\rvec-\rvec_n)\fvec_n .$
Evaluated at the stokeslet positions $\rvec=\rvec_m$, this gives the imposed velocities
$\uvec_m=\sum_{n\ne m}\Omat(\rvec_m-\rvec_n)\fvec_n .$
In addition, each stokeslet experiences a self-drag. Assigning a
hydrodynamic radius $a_{\rm eff}$ gives the self-contribution
\begin{equation}
\uvec_m^{\rm self}
=
\frac{1}{6\pi\mu a_{\rm eff}}\fvec_m .
\end{equation}
Together these relations give all $3N$ components of the imposed velocities $\{\uvec_m\}\equiv\uu$ as linear functions of the forces $\{\fvec_n\}\equiv\ff$:
\begin{equation}
\uu=\mathcal G\,\uline{\hbox{f}} .
\label{eq:grandMatrix}
\end{equation}
where 
\[  
     \mathcal{G}
=
\begin{pmatrix}
\Gmat_{11} & \cdots & \Gmat_{1N} \\
\vdots         & \ddots & \vdots           \\
\Gmat_{N 1} & \cdots & \Gmat_{N N}
\end{pmatrix}
\in \mathbb R^{3N \times 3N},
\]
and
\[
\begin{array}{cll}
    \Gmat_{mn} &= \Omat(\rvec_m - \rvec_n) &\hbox{for}~~ m\notequal n
    \\
    &= \frac{1}{6\pi\mu a_{\rm eff}} \Imat &\hbox{for}~~ m=n
\end{array}
\]

For rigid–body motion, the stokeslet velocities $\uvec_n$ are not independent. Each is determined by the translational velocity $\Vvec$ and angular velocity $\wvec$ of the body, defined with respect to a chosen origin:
\begin{equation}
\uvec_n = \Vvec + \wvec \times \rvec_n .
\end{equation}
Thus any rigid motion $(\Vvec,\wvec)$ corresponds to a $3N$-vector $\uu=\{\uvec_n\}$.  Knowing $\uu$ for six independent choices of $(\Vvec,\wvec)$ provides a basis from which $\uu$ for arbitrary $(\Vvec,\wvec)$ may be constructed by linear superposition.
For any imposed $\uu$, the corresponding forces $\ff$ follow from \eqref{eq:grandMatrix}.  
A unique solution exists because 
\hd{${\mathcal G}$ is symmetric positive-definite, implying $\uu\!\cdot\!\ff=\uu^{\mathsf T}{\cal G}\uu>0$ for any nonzero $\uu$ in Stokesian dynamics.}
The net force and torque exerted on the fluid are then $\Fvec=\sum_n \fvec_n$, $\tvec=\sum_n\rvec_n\times\fvec_n $. Evaluating these for the six basis motions yields the force and torque associated with arbitrary $(\Vvec,\wvec)$ by linearity.  For applications such as sedimentation, where force and torque are prescribed, we invert this relation to obtain the $6\times6$ mobility matrix defined by
\begin{equation}
\begin{pmatrix}
\Vvec\\[3pt]\wvec
\end{pmatrix}
=
\begin{pmatrix}
\Amat & \Bmat^{\mathsf T}\\[3pt]
\Bmat & \Cmat
\end{pmatrix}
\begin{pmatrix}
\Fvec\\[3pt]\tvec
\end{pmatrix}.
\end{equation}

This $6\times6$ matrix is denoted $\Mmat$. Like $\mathcal G$, 
\hd{it is symmetric and positive-definite in Stokesian dynamics, reflecting Onsager reciprocity and positive viscous dissipation.}
Although the solution of Eq.~\eqref{eq:grandMatrix} depends on the chosen stokeslet radius $a_{\rm eff}$, the resulting rigid-body mobilities should not. For sufficiently fine discretizations, we will show that taking $a_{\rm eff}=\beta\,\ell$, where $\beta\simeq0.25$ and $\ell$ is \hd{defined as the median of the nearest-neighbor distances between stokeslets}, balances self and interaction contributions and yields converged results largely independent of $a_{\rm eff}$. Further algorithmic and implementation details are provided below and in the Supplementary Information.

Given the rigid-body mobility tensor $\Mmat$, the sedimentation dynamics of a particle under constant external forcing can be formulated in both the laboratory and body frames. We take all external forces on the particle to act at a single point that is fixed with respect to the body, which we refer to as the \emph{force center}. All translational velocities $\Vvec$ and torques $\tvec$ are measured with respect to this force center. For gravitational forces, this point \wc{typically} 
coincides with the CoM, and there is no external torque on the particle. \wc{For nonuniform density (e.g., buoyant and heavy regions), the force center depends on the fluid density and thus generally differs from the CoM.}
The body frame is introduced because the mobility tensor is a geometric property of the particle and therefore remains fixed only in the frame attached to the body, whereas the applied forces, the particle orientation, and the resulting motion are most naturally described in the laboratory frame. 
We consider an applied force $\Fvec^{\mathrm{lab}}$ and torque $\tvec^{\mathrm{lab}}$ specified in the laboratory frame. 
The body frame differs from the lab frame by a translation $\chi(t)$, and a rotation $\Rotmat(t)$.  We take $\chi(t)$ to be the position of the force center in the lab frame at time $t$. The rotation matrix relates force and torque as seen in the lab frame to those seen in the body frame:
\begin{equation}
\Fvec^{\mathrm{body}}(t)=\Rotmat(t)^{\mathsf T}\Fvec^{\mathrm{lab}},
\qquad
\tvec^{\mathrm{body}}(t)=\Rotmat(t)^{\mathsf T}\tvec^{\mathrm{lab}}.
\end{equation}

The resulting translational and angular velocities, expressed in the body-frame basis, then follow as
\begin{equation}
\begin{pmatrix}
\Vvec^{\mathrm{body}}\\[3pt]\wvec^{\mathrm{body}}
\end{pmatrix}
=
\begin{pmatrix}
\Amat & \Bmat^{\mathsf T}\\[3pt]
\Bmat & \Cmat
\end{pmatrix}
\begin{pmatrix}
\Fvec^{\mathrm{body}}\\[3pt]\tvec^{\mathrm{body}}
\end{pmatrix},
\end{equation}

Note that $\Vvec^{\mathrm{body}}$ and $\wvec^{\mathrm{body}}$ represent the components of the instantaneous rigid-body motion expressed in the body-frame basis, rather than velocities measured relative to that frame. Mapping back to the laboratory frame is obtained via
\begin{equation}
\Vvec^{\mathrm{lab}}=\Rotmat(t)\,\Vvec^{\mathrm{body}},
\qquad
\wvec^{\mathrm{lab}}=\Rotmat(t)\,\wvec^{\mathrm{body}}.
\end{equation}

The force center therefore moves according to
\begin{equation}
\dot{\boldsymbol\chi}(t)
=
\Rotmat(t)\,\Amat\,\Rotmat(t)^{\mathsf T}\Fvec^{\mathrm{lab}}
\end{equation}
while the particle orientation dynamics follow from rigid-body kinematics,
\begin{equation}
\dot{\Rotmat}(t)
=
[\wvec^{\mathrm{lab}}(t)]_\times\,\Rotmat(t),
\end{equation}
where $[\wvec]_\times$ denotes the skew-symmetric matrix representing the cross product with $\wvec$. 
\hd{In the absence of torque,} $\wvec$ arises solely from $\Bmat$.
Writing the force as $\Fvec^{\mathrm{body}}(t)=F_0\,\hat{\Fvec}(t)$,
where $F_0=\|\Fvec^{\mathrm{body}}\|$ is constant,
we write the kinematic relation for a vector fixed in the laboratory frame,
\begin{equation}
\dot{\hat{\Fvec}}(t)
=
-\wvec^{\mathrm{body}}(t)\times\hat{\Fvec}(t)
=
- F_0 \big(\Bmat\,\hat{\Fvec}(t)\big)\times\hat{\Fvec}(t).
\label{eq:F_dynamics_dimensional}
\end{equation}
This preserves $\|\hat{\Fvec}\|=1$ and thus confines the dynamics to the unit sphere.
This formulation provides a compact geometric description of fixed points, limit cycles, and closed orbits in sedimentation trajectories \cite{witten2020review}. Further derivations and implementation details are given in the Supplementary Information.

\subsection*{Results}

To demonstrate the versatility of our STL-to-Stokeslet method, we compute the rigid-body mobility tensor and the resulting trajectories for a variety of complex particle shapes, as shown in Fig.~\ref{fig1}(a--d). For each particle, the geometry is imported from an STL file and remeshed to obtain a more uniform surface discretization. Prior to the mobility calculation, each particle geometry is uniformly rescaled to lie within an inscribing unit sphere. 
Consequently, sedimentation trajectories are expressed in the
dimensionless coordinates $(x^*,y^*,z^*)$, where lengths are
nondimensionalized by the radius of the particle’s circumscribed sphere.
For the DNA-shaped and chiral snowflake particles, the translation--rotation coupling tensor $\Bmat$ is symmetric when defined about the CoM. In these cases, the force direction on the unit sphere follows closed orbits, giving rise to complex CoM trajectories such as helices and spirograph-like patterns. Different initial orientations lead to distinct trajectory families, as shown in Fig.~\ref{fig1}(e--h). 

In contrast, both the non-chiral dolphin-like particle and the ice-grain-shaped one~\cite{nicolov2025dynamics} converge to fixed points of the force-direction dynamics on the unit sphere.
For the dolphin-like particle in Fig.~\ref{fig1}(c), the force direction rapidly converges to a stable fixed point given by $\Fvec_\star$. At this orientation, the angular velocity
is negligibly small, resulting in straight-line sedimentation without appreciable spin, as shown in Fig.~\ref{fig1}(g), analogous to that of a slender cylindrical rod with uneven mass distribution.
By contrast, for the ice-grain-shaped particle in Fig.~\ref{fig1}(d), the tensor $\Bmat$ contains a large antisymmetric component,  implying a chiral response~\cite{krapf2009}. Although the force direction again converges to a fixed point, the resulting CoM trajectory is helical, as shown in Fig.~\ref{fig1}(h). These two examples illustrate that fixed points of the force-direction dynamics in the body frame can produce qualitatively different motions in the laboratory frame. Specifically, 
the translational and angular velocities are given by
$\Vvec^{\mathrm{body}}=\Amat\,\Fvec_\star$ and
$\wvec^{\mathrm{body}}=\Bmat\,\Fvec_\star$.
The relative orientation of $\Vvec^{\mathrm{body}}$ and $\wvec^{\mathrm{body}}$ determines the observed trajectory: straight-line sedimentation occurs when $\wvec^{\mathrm{body}}$ is negligible or when $\Vvec^{\mathrm{body}}\parallel\wvec^{\mathrm{body}}$, whereas non-collinearity produces helical motion.


\begin{figure}[!htbp]
\centering
\includegraphics[width=\linewidth]{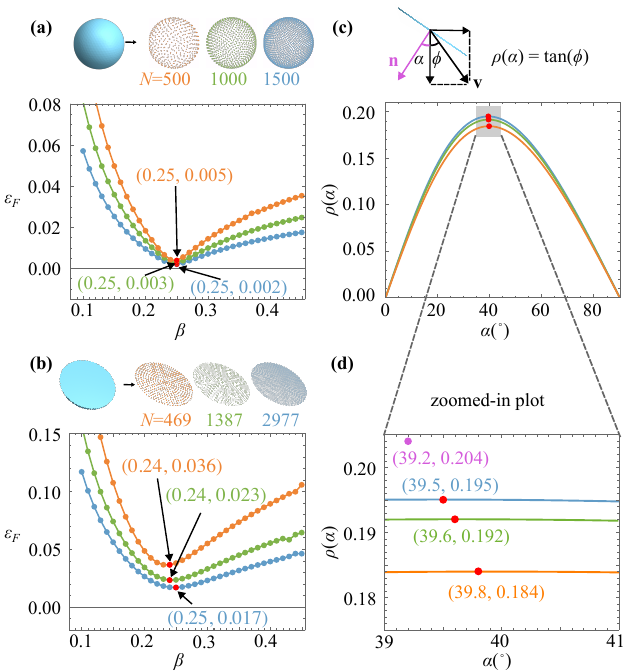}
\caption{
\textbf{Validation of the STL-to-Stokeslet method for a sphere and a thin disk.}
\textbf{(a)} Sphere discretized with increasing numbers of stokeslets. 
The mobility error is defined in Eq.~\ref{error} and is evaluated as a function of the stokeslet radius factor $\beta$. 
As the number of stokeslets increases, the error decreases and consistently exhibits a minimum near $\beta=0.25$.
\textbf{(b)} Thin circular disk discretized with increasing numbers of Stokeslets. 
As in the spherical case, the mobility error decreases with increasing $N$.
\textbf{(c)} Analytical prediction of the horizontal-to-vertical velocity ratio, $\rho(\alpha)$, for a thin disk as a function of the tilt angle $\alpha$. 
The inset illustrates the geometric definition of $\alpha$ and the decomposition of the sedimentation velocity into vertical and horizontal components.
\textbf{(d)} Enlarged view of the highlighted region in \textbf{(c)} near the maximum of  $\rho(\alpha)$. 
As $N$ increases, $\alpha$ and $\rho(\alpha)$ converge to the analytical predictions ($\alpha^\ast = 39.2^\circ$, $\rho^\ast(\alpha^\ast)=0.204$) \cite{happel2012low}.
}
\label{fig2}
\end{figure}

To validate our computational method, we first apply it to a sphere of radius $a$, compute the corresponding $6\times6$ mobility tensor, and compare the result with the analytical expression~\cite{happel2012low},
\begin{equation}
\Mmat^\ast_{\rm sphere} =
\begin{pmatrix}
\frac{1}{6\pi\mu a}\,\Imat_3 & \mathbf 0\\[4pt]
\mathbf 0 & \frac{1}{8\pi\mu a^3}\,\Imat_3
\label{sphere}
\end{pmatrix}.
\end{equation}
Denoting the numerical mobility tensor by $\widetilde{\Mmat}$, we quantify its deviation from the mobility of the analytical sphere $\Mmat^\ast$ using the normalized Frobenius-norm error, a standard measure of matrix distance that compares all components on equal footing.
.
\begin{equation}
\varepsilon_F
=
\frac{\|\widetilde{\Mmat}-\Mmat^\ast\|_{\mathrm F}}{\|\Mmat^\ast\|_{\mathrm F}}
=
\left(
\frac{\sum_{i,j}
\left|\widetilde{M}_{ij}-M^\ast_{ij}\right|^2}{\sum_{i,j}
\left|M^\ast_{ij}\right|^2}
\right)^{1/2}.
\label{error}
\end{equation}
For a range of Stokeslet radius factors~$\beta$, we evaluate the error $\varepsilon_F(\beta)$. As the number of Stokeslets $N$ increases, the error decreases and consistently approaches a minimum near $\beta=0.25$, as shown in Fig.~\ref{fig2}(a). This behavior indicates that $\beta\approx0.25$ provides a robust and near-optimal choice across different discretization levels $N$.

We also validate the method using a thin circular disk of radius $a$ and thickness $h\!\ll\!a$. 
Unlike a sphere, the translational block of the analytical mobility tensor for a disk is anisotropic in the body frame~\cite{happel2012low,kanwal1970note,zhang1998oscillatory},
\begin{equation}
\Mmat^\ast_{\rm disk}
=
\begin{pmatrix}
\tfrac{1}{32\mu a} \mathrm{diag}\!\left(3,3,2\right) & \mathbf 0\\
\mathbf 0 &
\tfrac{3}{32\mu a^3}\,\Imat
\label{disk}
\end{pmatrix}.
\end{equation}
We use this analytical mobility tensor in Eq.~\ref{error} to assess the convergence of the numerical results. As shown in Fig.~\ref{fig2}(b), the error decreases systematically with increasing $N$, indicating convergence with respect to the surface discretization. 
\wc{For objects with dense, uniform meshes (like the sphere and disk), the convergence of both $\Amat$ and $\Cmat$ is logarithmic (see Fig.~S2). However, we expect the rotational response (the $\Cmat$ block) to dominate the error for more general meshes since torques are sensitive to the placement of discrete stokeslets, and the hydrodynamic screening by the stokeslets is incomplete.}


Further, the optimal Stokeslet radius factor is found to be $\beta\simeq 0.25$, consistent with the behavior observed for spherical geometries. The motion of a thin circular disk under a constant external force follows a straight-line trajectory, consisting of both vertical settling and horizontal drift components. The direction of this trajectory depends on the initial orientation of the disk: different tilt angles, $\alpha$, give rise to different drift angles, $\phi$. 
As illustrated in Fig.~\ref{fig2}(c,d), we extract $\phi$ by varying $\alpha$ over the range $0^\circ$--$90^\circ$. 
For each value of $\alpha$, we compute the ratio of the horizontal to vertical components of the translational velocity $\rho(\alpha)=\tan(\phi)$.
The maximum of this curve occurs at an angle $\widetilde{\alpha}$, with a corresponding value $\rho(\widetilde{\alpha})$. These numerical results are compared with the analytical predictions for a thin disk, namely the tilt angle $\alpha^\ast = 39.2^\circ$ yielding the maximum velocity ratio $\rho^\ast(\alpha^\ast)=0.204$ \cite{happel2012low}. 
As $N$ increases, both $\widetilde{\alpha}$ and $\rho(\widetilde{\alpha})$ converge to their analytical values.

Moreover, both the spherical and thin-disk validations exhibit an optimal stokeslet radius factor that converges to $\beta \simeq 0.25$ when minimizing the Frobenius-norm–normalized error, $\varepsilon_F$. 
This behavior can be understood from the structure of the singular self-interaction term $\mathbf G_{nn}$ in our STL-to-Stokeslet method, which is controlled by an isotropic self term proportional to $(\beta\, \ell)^{-1}$. 
As shown in the Supplementary Information, analysis of the local Stokeslet surface integral in the dense-mesh limit reveals that the dominant contribution arises from a surface patch with characteristic length scale $\sim \ell$, leading to a self-mobility scaling $\mathbf G_{nn}\sim C/(\mu \ell)$. 
The dimensionless prefactor $C$ depends only on the local discretization geometry and is independent of the total number of surface nodes $N$. To estimate this prefactor, we approximate the local surface patch by a flat disk and explicitly integrate the stokeslet kernel, which yields an anisotropic local response
$\mathbf G_{\mathrm{loc}}=\frac{1}{8\pi\mu a}\,\mathrm{diag}(3,3,2)$,
with larger tangential than normal components. 
We approximate this anisotropic response with an isotropic scalar self term, $\mathbf G_{\mathrm{loc}} \approx (\alpha^\star/8\pi\mu a)\Imat$, and determine the optimal prefactor $\alpha^\star = 31/11$
by minimizing a weighted least-squares mismatch between tangential and normal components of the local mobility tensor
$2w_t(\alpha-3)^2 + w_n(\alpha-2)^2$,
with weights $w_t$ and $w_n$ determined by their relative contributions across the six rigid-body motions.

Matching this effective isotropic response to the discrete self term then yields $\beta \simeq 0.25$, in quantitative agreement with the numerically observed optimum for dense, quasi-uniform discretizations.

\begin{figure}[!htbp]
\centering
\includegraphics[width=\linewidth]{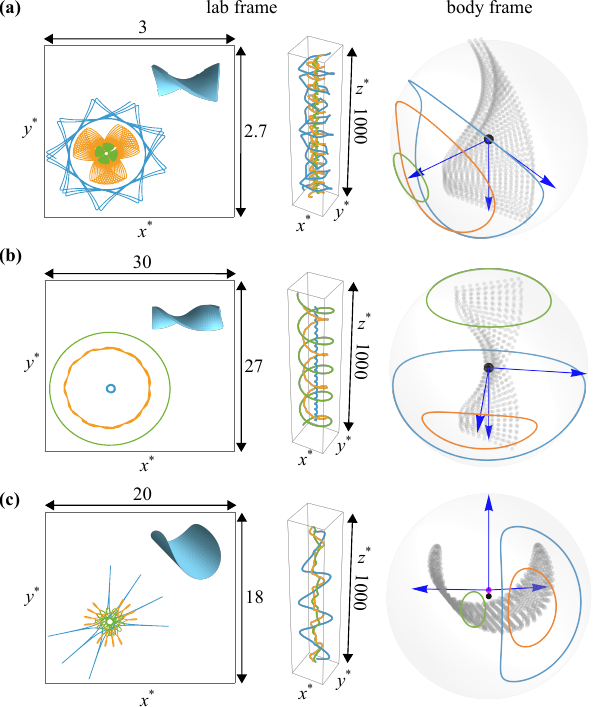}
\caption{
\textbf{Validation of the STL-to-Stokeslet method for complex geometries.}
\textbf{(a--c)} Comparison of our computational results with reference sedimentation trajectories reported in the recent literature for three particles: helical ribbons of different lengths and a U-shaped disk.
\textbf{Left panels:} Predicted CoM trajectories in the $x$ -- $y$ plane for different initial orientations.
\textbf{Middle panels:} 3D sedimentation trajectories for the same particles.
\textbf{Right panels:} Corresponding force-direction trajectories on the unit sphere. Magenta and black points denote the CoM and CoR, respectively.
}
\label{fig3}
\end{figure}

In addition to these validation shapes, we further test our method by applying it to complex particle geometries recently studied in the literature: helical ribbons~\cite{huseby2025helical} and U-shaped disks~\cite{miara2024dynamics}.
These geometries lack analytical mobility tensors and therefore provide a challenging extension for assessing whether our approach can accurately reproduce the sedimentation dynamics of arbitrary 3D shapes. Figure~\ref{fig3}(a,b) show trajectories for helical ribbons with twist angles $3\pi/4$ and $4\pi/3$, respectively, in agreement with Fig.~5 of Ref.~\cite{huseby2025helical}. For the U-shaped disk, the trajectories shown in Fig.~\ref{fig3}(c) reproduce the dynamics reported in Fig.~6(b--d) of Ref.~\cite{miara2024dynamics}.  In both cases, our results show excellent qualitative agreement with recent experimental measurements. In particular, the computed motions reproduce the spirograph-like trajectories observed for both shapes, and accurately reproduce the large-amplitude superhelical motion of the ribbon as well as the characteristic pitching and rolling dynamics of the U-shaped disk. This agreement demonstrates that the mobility tensors generated by our method faithfully capture the hydrodynamic response of complex particle shapes.


\begin{figure}[!htbp]
\centering
\includegraphics[width=\linewidth]{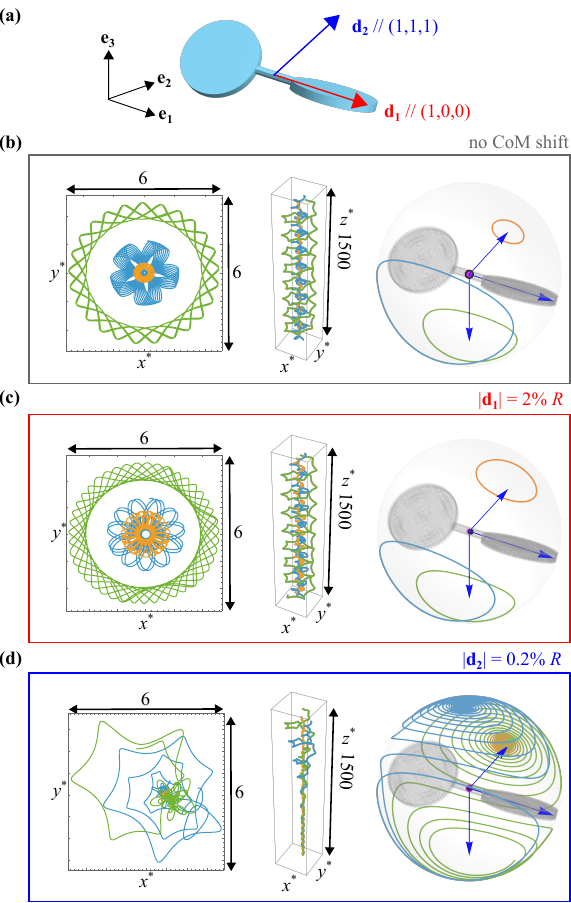}
\caption{
\textbf{Effect of CoM shift on sedimentation dynamics.}
\textbf{(a)} Schematic illustration of a double-paddle particle and its principal orthogonal body axes. $(\mathbf e_1,\mathbf e_2,\mathbf e_3)$.  A CoM shift is introduced by displacing the mass center either along the body axis $\mathbf e_1$ ($\dvec_1$) or along a direction not aligned with any principal body axis ($\dvec_2$).
\textbf{(b)} 
Sedimentation dynamics of the particle with no CoM shift.
\textbf{Left:} CoM trajectories in the $x^*$--$y^*$ plane.
\textbf{Middle:} 3D sedimentation trajectories.
\textbf{Right:} Force-direction trajectories on the unit sphere.
The unshifted particle exhibits closed orbits, consistent with chiral sedimentation behavior.
\textbf{(c)} Sedimentation dynamics for CoM shifts along $\dvec_1$, with magnitude $|\dvec_1| = 2\%\, R$, where $R$ is the radius vector of the circumscribed sphere of the particle. 
\textbf{(d)} Sedimentation dynamics for CoM shifts along $\dvec_2$.
Even a $0.2\%$ offset dramatically changes the trajectories, producing spiraling trajectories and the force direction paths converge to two fixed points on the unit sphere. Magenta and black points denote the CoM and CoR, respectively.
}
\label{fig4}
\end{figure}

In nature, particulates are often composed of multiple materials or minerals. The fabrication of particles can also introduce unavoidable density inhomogeneities~\cite{huseby2025helical}, resulting in shifts of the CoM that are difficult to control or detect directly. 
Fortunately, the $\Amat$, $\Bmat$, and $\Cmat$ matrices are sufficient to determine how a shift of the force center modifies the mobility matrix, and hence the resulting motion \cite{happel2012low}.  
Using the translation theorems for mobility tensors \cite{kim2013microhydrodynamics}, 
the translation--rotation coupling becomes
\begin{equation}
\Bmat' = \Bmat + \Cmat\,[\dvec]_\times\,,
\label{eq:B_shift}
\end{equation}
where \hd{$\dvec$ is the vector of CoM shift/displacement, and} $[\dvec]_\times$ maps a force to the corresponding torque induced by the CoM shift. Equation~\eqref{eq:B_shift} shows that a CoM shift generically introduces an antisymmetric contribution to $\Bmat$, which underlies the qualitative changes in the resulting sedimentation dynamics. See the Supplementary Information for further derivation details. \wc{The numerical values of $\widetilde{\Mmat}$ for the objects in Figs.~\ref{fig3} and ~\ref{fig4} are listed in the Supplementary Information to facilitate comparison with other calculations.}

As shown in Fig.~\ref{fig4}, the sedimentation behavior of a double-paddle exhibits distinct sensitivity to the direction of the CoM shift. 
\wc{By symmetry, the CoR of this object lies at its geometric center. When the CoM coincides with this point, }
the coupling tensor $\Bmat$ is symmetric, and the particle exhibits periodic spirograph-like trajectories in the laboratory frame together with closed orbits on the unit sphere in the body frame,  as studied numerically in \cite{makino2003}. 
Introducing a small CoM shift of $|\dvec_1| = 2\% \,R$ ($R$ being the radius of the circumscribed sphere of the particle) along the long principal body axis, $\mathbf e_1$, as shown in Fig.~\ref{fig4}(c), $\Bmat$ remains symmetric, leading to modified but still periodic trajectories. 
In contrast, when a CoM shift is applied along a direction $\dvec_2$ that is not aligned with any principal body axis, even a small offset of $0.2\%$ introduces a substantial antisymmetric component into $\Bmat$. 
This symmetry breaking produces spiraling CoM trajectories and two fixed points on the unit sphere, as shown in Fig.~\ref{fig4}(d).

Besides diagnosing the CoM of the fabricated particles, these results demonstrate that CoM shifts act as a powerful control parameter for sedimentation behavior.
Small CoM shifts along the long principal body axis primarily lead to quantitative modifications of the motion, whereas a tiny generic off-axis shifts qualitatively reorganize the structure of the mobility tensor and readily induce fixed points that govern the resulting alignment behavior.
The explicit $\Amat$, $\Bmat$, and $\Cmat$ matrices obtained from the stokeslet method provide the means to predict the wide range of motions accessible through shifts of the force center. Thus capturing the transition from periodic or quasiperidoic motion to dynamics featuring spiraling trajectories toward fixed points highlights its utility for predicting, diagnosing, and designing sedimentation behavior for complex-shaped particles.

\subsection*{Discussion}

In summary, we have developed an STL-to-Stokeslet framework that enables the direct computation of the rigid-body mobility tensor for arbitrarily shaped particles from their STL file. By discretizing the continuous surface traction with Stokeslets, the method avoids geometry-specific modeling assumptions and does not rely on experimental calibration. It converges systematically under refinement of the surface discretization and is therefore broadly applicable to rigid bodies of complex 3D shape. The accuracy and robustness of the method were established through benchmark validations against spherical and thin-disk geometries with known analytical results. 
We then demonstrated its generality by qualitatively reproducing previously reported sedimentation dynamics for helical ribbons and U-shaped disks, two nontrivial particle geometries that lack analytical mobility tensors.

Beyond computing the mobility tensor and the resulting dynamics, the framework naturally incorporates shifts of the CoM relative to the geometric centroid through corresponding modifications of the mobility tensor. This unified treatment allows CoM offsets to be analyzed quantitatively without introducing additional modeling assumptions or recomputing the mobility matrix. 
Our results demonstrate that even CoM shifts of less than 1\% can qualitatively alter sedimentation dynamics, inducing fixed points on the unit sphere and breaking periodic sedimentation into spiraling motion, thereby revealing a pronounced sensitivity to off-axis perturbations. As such, CoM shifts emerge as a powerful and experimentally accessible control parameter for diagnosing and tuning sedimentation behavior across different particle geometries.

\wc{Despite the versatility of our method, there are limitations. The stokeslet discretization does not perfectly enforce a rigid, no-slip boundary. Because the velocity field imposed by discrete stokeslets only approximately matches the rigid-body motion, the fluid inside the object exhibits a small residual flow, reflecting imperfect hydrodynamic screening~\cite{witten2020review}. However, for many objects of interest, the assumption of a perfectly rigid body with a smooth no-slip surface is itself an idealization. Quantities such as the effective hydrodynamic radius can be sensitive to these modeling assumptions. Our method is therefore not intended for situations in which the surface geometry is known with high accuracy and very high numerical precision is required. 
In addition, high precision is intrinsically difficult to attain within the present approach. First, the numerical error decreases only slowly with the number of stokeslets $N$, with convergence that is approximately logarithmic (see Supplementary Information). Second, the computational cost scales as $O(N^2)$, so improving precision by increasing $N$ rapidly becomes expensive. \hd{Further refinements of the model, such as the replacement of the Kirkwood-Riseman mobility by the Rotne-Prager one \cite{rotne1969}, or the use of regularized stokeslets \cite{cortez2005},} \jb{
may be necessary for specialized cases.}}

Finally, while the present work focuses on rigid particles, the STL-to-Stokeslet framework naturally extends to weakly deformable bodies. In the limit of small elastic deformations, shape changes and the additional forces they introduce can be treated as perturbations about a reference rigid geometry, leading to systematic corrections to the hydrodynamic mobility within the same stokeslet-based formulation. 
Together with implementations in both \textit{Mathematica} and \textit{Python}, this flexibility positions the method as a reproducible and extensible platform for studying, predicting, and diagnosing the dynamics of complex particles at low-Reynolds-number.

\subsection*{Materials and Methods}

Although the present method is designed as a simpler alternative to more cumbersome numerical techniques, its accuracy can be assessed by comparison with established approaches such as embedded-surface or boundary-integral formulations. These methods provide a useful reference standard for geometries lacking analytic solutions.

The steps below describe a Mathematica-based implementation, but the procedure itself is not software-specific. 
The STL geometry is imported into Mathematica and converted into a surface mesh object. 
Equivalent STL-to-mesh and boundary representations can also be constructed in Jupyter notebooks using standard Python libraries (e.g.\ \texttt{trimesh}, \texttt{numpy-stl}).
Particle geometries were imported from STL files as \texttt{BoundaryMeshRegion} objects, yielding an explicit triangulated representation of the particle surface on which all subsequent mesh processing and discretization steps were performed. To improve numerical stability and control discretization quality, surface meshes were processed using built-in mesh operations: \texttt{Remesh} was employed to generate quasi-uniform triangulations when starting from highly nonuniform STL meshes, while \texttt{SimplifyMesh} was used to reduce the number of surface nodes without altering the overall particle shape. The resulting surface nodes $\{\rvec_n\}_{n=1}^{N}$ were expressed relative to the geometric center of the particle shape, computed using \texttt{RegionCentroid}, which was taken as the reference origin unless otherwise specified. All geometries were uniformly rescaled to lie within an inscribing unit sphere. Typical discretizations employed $N \sim 2\times10^3$--$2\times10^4$ surface points, with larger values used for geometries with fine features.

The grand mobility matrix $\mathcal{G} \in \mathbb{R}^{3N \times 3N}$ was assembled explicitly as a dense matrix in double precision (\texttt{Real64}), and all linear systems
$\uu = \mathcal{G}\,\ff$
were solved using a direct dense solver. In \textit{Mathematica}, we constructed a reusable factorization via \texttt{LinearSolve} applied once to $ G$, and subsequently solved for multiple right-hand sides corresponding to the six imposed rigid-body motions. This strategy avoids explicit matrix inversion while enabling efficient and numerically stable repeated solves with the same coefficient matrix.

\subsection*{Data, Materials, and Software Availability.}
The data and code used in this study, including the \textit{Mathematica} implementation, have been deposited in the Zenodo database (\url{https://doi.org/10.5281/zenodo.18761908}). The \textit{Python} version of the code is available on GitHub (\url{https://github.com/jcburtonlab/STL-to-Stokeslet-Method}).

\subsection*{Acknowledgments}
We thank Greg Voth for insightful discussions, and Wentao Yu for providing a 3D rendering of a microscale ice grain. This work was supported by the United States–Israel Binational Science Foundation (BSF) under Grant No. 2022229, and by the Gordon and Betty Moore Foundation, Grant DOI: 10.37807/gbmf12256. We also acknowledge support from the Laney Graduate School SOAR program at Emory University.  

\bibliographystyle{unsrt}
\bibliography{references}

\end{document}